# The typical manners of dynamic crack propagation along the metal/ceramics interfaces: A molecular dynamics study

Yanguang Zhou[1], Zhenyu Yang[1*], Tao Wang[2], Dayong Hu[3], Xiaobing Ma[4]

[1]Institute of Solid Mechanics, Beihang University (BUAA), Beijing 100191, P. R. China

[2]Interdisciplinary Centre for Advanced Materials Simulation (ICAMS) Ruhr-Universität Bochum

[3]School of Transportation Science and Engineering, Beihang University (BUAA), Beijing 100191, P. R. China

[4]Institute of reliability engineering , Beihang University (BUAA), Beijing 100191, P. R. China

*Corresponding author E-mail: zyyang@buaa.edu.cn (Z. Y. Yang)

Abstract

This letter addresses the issue of interfacial crack propagation mechanisms on various interfaces with using molecular dynamics (MD). Four different interfacial crack propagation manners are recognized by MD simulations: (1) the crack propagates along the interface strictly; (2) the nucleation of a twinning at the crack tip leads to the crack tip blunting; (3) the daughter crack appears ahead of the main crack and then coalesces to the mother crack; (4) the crack tip blunts with stacking fault nucleation at the crack tip. Furthermore, the adhesive strength coefficient $\lambda$ is used to identify the type of the interface. Interface with $\lambda > 0.9$ means an "ordered interface", and interfacial crack propagates in mode I or mode II. While $\lambda < 0.9$ represents a "disordered interface", and mode III and mode IV of the interfacial cracks are preferred. This work illustrates the effect of adhesive strength of interface on the mechanism of crack propagation along the interface of metal/ceramics nanocomposites.

Nano materials (NMs) have gained a vast majority of attention in recent years due to its unique mechanical, electrical and optical properties[1-4]. As the dimensions decrease to nanoscale, interfaces become a significant important part to influence the characters of the NMs[5,6]. Thus, it is necessary to investigate the fracture behaviors of the interface, which is extremely helpful for understanding the failure mechanism of NMs.

Many researches have demonstrated, through experiments, that cracks in the interfacial region can propagate in both the brittle and ductile manners[7-12]. Simulations reported in the previous literatures showed that the propagation of interfacial cracks can be affected by the external temperature[13] and loading angle[14,15]. Meanwhile, the anisotropic lattice of the crystalline metal phase can change the style of crack propagation (from brittle to ductile) as well[13,16-18]. Because of the adhesive strength of interface show significant effects on the interfacial properties, it is of great interest to study its roles in the interfacial crack propagation. In the single phase material, the crack can propagate in a strict brittle or ductile manner[19-22]. However, the interfacial region of metal/ceramic composite is influenced by both the metal and the ceramic phase[14,23,24]. Thus, crack propagation in the interfacial region can be quite different from that in the single phase materials. Until now, there are still no studies about how the adhesive strength of interface affects the crack propagation behaviors in the interfacial region.

In this letter, molecular dynamics (MD) simulations are performed to investigate the effects of adhesive strength on the crack propagation at the SiC/Al interface. We first describe the details of the simulation method: the atomistic potential, the setup of the model, the defect identification and the temperature control. The interatomic potentials of Al and SiC are in the form of embedded atom method[25] and the tersoff method[26], respectively. In addition, morse potential is adopted to depict the interaction between the Al and SiC, which s fitted from the *ab initio* values[23,27]. These potentials have been proved to describe the interactions among atoms well[23]. The simulation cells are shown in Fig. 1. In all the cases, the systems have a planar structure with all metal phases having a common [110] (defined as the z direction), which allows two potential active (111) slip systems in faced-center-cubic (FCC) metals. The size of the cells is about 85 nm × 55 nm × 2 nm. Periodic boundary conditions are implemented along X and Z directions. A 5 nm crack in the system is nucleated by excluding the atomic interactions between atoms at both sides of the interface of Al and SiC. The central symmetrical parameter (CSP) is used to classify the defective atoms in the simulated cells. They are atoms with the CSP value of 0 arranged in FCC structure, those with CSP values of 0.5~1.25 and 4.0~6.0 are dislocation cores and stacking faults, respectively. The CSP value of the free surface atoms is above 23.0. After the initial construction, conjugate gradient method is used to minimize the whole system to obtain equilibrium configurations. Then the system is thermally equilibrated to approximately 0 K for 50 ps with using a NVT ensemble. Starting from the equilibrium configuration of the system, an uniaxial strain is applied with a strain rate of $10^9 \, s^{-1}$ along the Y direction.

Eight interface models with metal phases in various crystal orientationsand the ceramic phase in a constant crystal orientation are investigated in this paper. Table 1 shows the detail information of the models. As we know, the crack propagates along the interface can either in a brittle or a ductile mode, which is mainly determined by the interfacial atomic structures. Here, four crack propagation modes are observed in the simulations: (1) the crack propagates along the interface strictly in a brittle mode (Fig. 2a); (2) the nucleation of a twinning at the crack tip and then the crack tip blunts (Fig. 2b); (3) the daughter crack ahead of the mother crack appears and then coalesces to the mother crack, leading to the crack propagates in a brittle mode (Fig. 2c); (4) the crack tip blunts by stacking fault nucleation at the crack tip (Fig. 2d). To present the details of crack propagation, the four crack propagation modes mentioned above are defined as mode I, mode II, mode III and mode IV, respectively. For mode I, the crack propagates along the interface strictly (Fig. 2), which was also observed by Yang *et al.*[14] and Zhou *et al.*[28,29]. The crack tips (A in Fig. 3) are the only places of stress concentration, and the typical stress field as predicted by the theory appears at the crack tip. As Yang *et al.*[14] mentioned, the atomic configuration at the interface is regular for such kind of interface ("ordered interface" as defined in this paper later), then the influence of lattice mismatch between two phases can be ignored. Thus the main factor that affects the stress distribution is the far-field stress, which leads to the "butterfly shape" stress distribution at the crack tip, as shown in Fig. 3. While for the mode II, a leading Shockley partial dislocation is emitted at the crack tip (Fig. 4a) firstly. In Fig. 4b, the leading partial dislocation travels to a stable position at T = 25 ps, then the nucleation of a twinning (① in Fig. 4b) appears at the left side of the dislocation. The twin is only one atomic layer thick at the crack tip area and 8.1 nm long, which is thought to be initial width and length of the twin. At T = 29 ps, the second twinning arises at the right side of the dislocation (② in Fig. 4c) and the twin is two atomic layers thick at the base with a length of 12.5 nm. With the increase of the load, the twin width and length increase (Fig. 4d). What is also worth to be noticed is that the distance between two dislocation cores that far from the crack tip is nearly the same, which was also observed by Tadmor et al.[30,31]. In addition, the appearance of the twins is because of the high loads and the high strain rates ($> 10^7$/s). Otherwise the accompanying dislocation after the leading partial dislocation is the trialing dislocation, which was verified by Warner et al.[32].

When it comes to mode III, the crack tip is blunted when it extends to the disordered area which results from the lattice mismatch between the metal phase and the ceramic phase (Fig. 5a). Then the daughter crack ahead of the mother crack appears with the increasing loads (Fig. 5b). The daughter crack coalesces to the mother crack, leading to the extension of the crack finally (Fig. 5c). The formative crack surface is a rough surface rather than a glossy one which appears in the cases in the mode I (Fig. 3). In addition, if the shear stress in the slip plane is over the threshold value (in this case: 8.78 GPa for $(\bar{1}11)$ plane and 4.18 GPa for $(1\bar{1}1)$ plane) needing to activate the dislocations, the crack propagates with dislocations emitting from the tips (Fig. 5b and 5c). However, when the crack propagates in mode IV, the edge dislocation with a Burgers vector of $\frac{1}{6}[1\bar{1}2]$ emits from the crack tip (Fig. 6a), which can be viewed as the leading partial dislocation of a full dislocation with a burgers vector of $\frac{1}{2}[011]$. With the increase of the load, the atoms in the crack tip region become disorder and another edge dislocation emits at the crack tip (Fig. 6b). More dislocations appear in the disordered crack tip region with the loads increasing (Fig. 6c) and the stacking fault is generated finally (Fig. 6d). It is thought that the strong interactions between the two phases leads to local atom disorder in the crack tip, and it should be responsible to the phenomenon mentioned above[14,33].

In order to disclose the relationship between the crack propagation and the properties of interface, models (Fig. 7a and Fig. 7b) with the same orientations as mentioned in the crack models are loaded till entire fracture. It is interesting to find that the systems are broken in two manners: (1) perfect cleavage along the interface (Fig. 7c) and the atoms in interface are regular, we name this kind of interface "ordered interface" (OI); (2) rough Al surface with Al atoms sticking to the SiC surface (Fig. 7d) and atoms in interface are disordered, such interfaces are called "disordered interface" (DI). Thus the actual energy $E_{act}$ needed to separate the interface of metal and ceramic can be calculated as

$$E_{act} = E_{t\_final} - E_{t\_perfect} \tag{1}$$

where, $E_{t\_final}$ is the total energy of a sample when the system is separated sufficiently (Fig. 7c and Fig. 7d), which ensure that two slabs of interface do not interact. $E_{t\_perfect}$ is the total energy of an equilibrium system (Fig. 7a and Fig. 7b).

Meanwhile, the theoretical energy $E_{theo}$ that needs to separate the two parts of the system is calculated as well, which is regarded as the energy needed to divide the

two parts along the interface strictly, and can be expressed as

$$E_{theo} = E_{t\_separate} - E_{t\_perfect} \qquad (2)$$

where, $E_{t\_separate}$ is the total energy of a stable sample consisting of two slabs which are separated far enough to ensure that two slabs of interface do not interact with each other.

Then, an adhesive strength coefficient (ASC) $\lambda$ can be defined to identify the characters of interfaces,

$$\lambda = \frac{E_{act}}{E_{theo}} \qquad (3)$$

$E_{act}$, $E_{theo}$ and $\lambda$ of all models are listed in Table 2. Here, the interfaces (B, D and H) with the values of $\lambda$ over 0.9 are found to be the OIs, while other interfaces with $\lambda$ between 0.8 and 0.9 (A, C, E, F and G) are observed to have the same properties of DIs. From the analysis above, we can find that crack propagations of mode I or mode II appear on the OIs. While for DIs, crack propagates along the interface with mode III and mode IV.

In summary, atomistic simulations are performed to investigate the dynamic crack propagation along Al/SiC interfaces. Four crack propagation modes are observed: (1) the crack propagates along the interface strictly with a brittle mode; (2) the nucleation of a twinning partial dislocation appears on the crack tip and the crack tip blunts; (3) the daughter crack appears ahead of the mother crack and coalesces to the mother crack leading to the crack propagates in a brittle mode; (4) the crack tip blunts with stacking fault nucleation at the crack tip. The ASC $\lambda$ is used to distinguish the physical characteristic of interfaces, and the results show that $\lambda > 0.9$ for the OIs and $\lambda < 0.9$ for the DIs. Based on the simulated results, it is interesting to find that the ASC plays a vital role in the dynamic interfacial crack propagation. The simulations show the followings: (1) For the OIs with $\lambda > 0.9$, crack propagates along the interface with mode I or mode II. (2) While for the DIs with $\lambda < 0.9$, mode III or mode IV is observed for the interfacial crack propagations. This letter demonstrates that the modes of the interfacial crack propagation can be changed by the ASC dramatically.

The authors thank the support from the National Natural Science Foundation of China (Nos. 11002011, 11172027, 61104133 and 11102017) and the Fundamental Research Funds for the Central Universities and the Beijing Higher Education Young Elite Teacher Project.

**Figure captions**

FIG. 1. Diagram of the 3D Al/SiC computational cell with crack in the central. The blue regions are the loading layers.

FIG. 2. Four crack propagation modes are observed. (a) the crack propagates along the interface strictly (mode I); (b) the nucleation of a twinning partial dislocation appears on the crack tip and the crack tip blunts (mode II); (c) the micro crack ahead of the main crack generates and coalesces to the main crack leading to the crack propagates in a brittle mode (mode III); (d) the crack tip blunts with stacking fault forming at the crack tip region (IV). Central symmetry parameter (CSP) is used to identify the defective atoms, with FCC atoms in shallow blue, what should also be noticed is that the CSP of the ceramic atom is always 0 due to its stable structure. The red arrows in all the pictures represent the direction of crack propagation.

FIG. 3. Contour plots of the atoms shear stress field and crack propagation states of the pure brittle crack propagation under mode I loading at strains ($\gamma$) of (a) $\gamma = 0.030$, (b) $\gamma = 0.037$, (c) $\gamma = 0.048$. A and B represent the crack tips during the period of crack propagation. The maximum and minimum values of the shear stress (Sxy) are -3.55GPa (blue) and 3.55Gpa, (red) respectively. The red arrows stand for the direction of the crack propagation.

FIG. 4. MD snapshots of the asymmetrical crack propagation (including pure brittle and pure ductile manners) under model I loading. ① the first twin boundary ② the second twin grain boundary, and so on, ⑦ represents the seventh twin grain

boundary. (a) t = 25ps. (b) t = 27ps. (c) t = 30ps. (d) t = 39ps. The atoms are identified as in Fig. 2.

FIG. 5. The process of the quasi brittle manner of the interfacial crack.. Stress concentration appears at the crack tip and some places of the interfaces due to the strong lattice mismatch effect between the two phases. (a) t = 29ps. (b) t = 38ps. (c) t = 47ps. The atoms are colored by the same method in Fig. 3, while the magnitude of the shear stress varies from -4.8 GPa to 4.9 GPa.

FIG. 6. A series of snapshots monitoring the quasi ductile manner of the interfacial crack. The atoms are identified as in Fig. 2. (a) $\gamma = 0.029$, (b) $\gamma = 0.032$, (c) $\gamma = 0.037$. The pictures in the bottom right are the amplification of the region in the red circle.

FIG. 7. The sketch pictures of two typical fracture manners of the Al/SiC systems. (a) to (c): the cleavage mode of the interface; (b)-(d): the decohesion rupture of the Al/SiC model. (a) and (c) are the initial models after equilibrium. (b) the model after the fracture of cleavage. (d) the model after decohesion rupture.

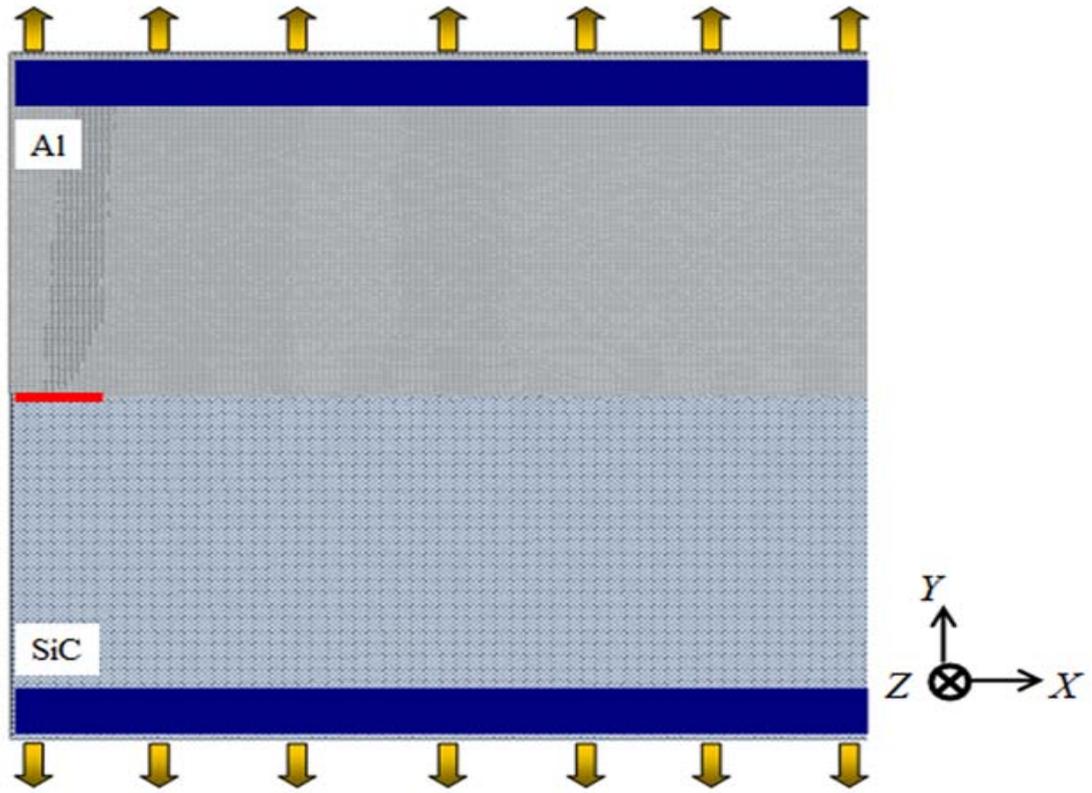

Fig. 1

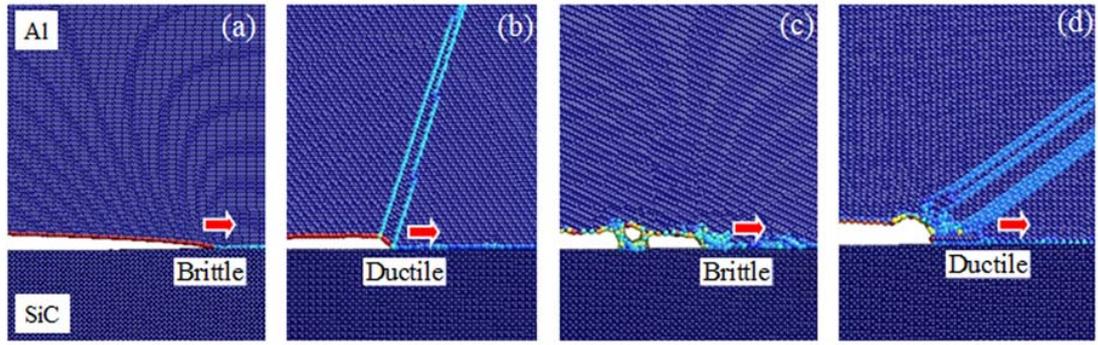

Fig. 2

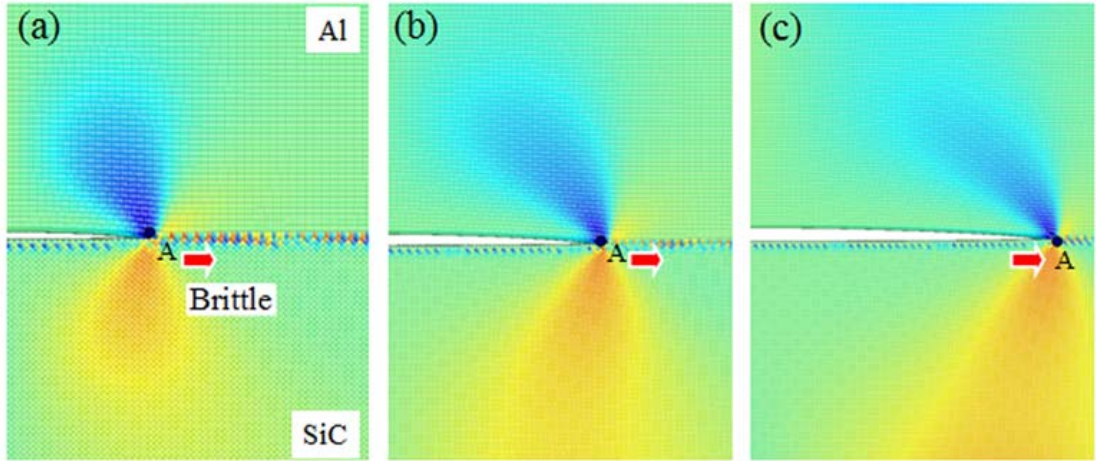

Fig. 3

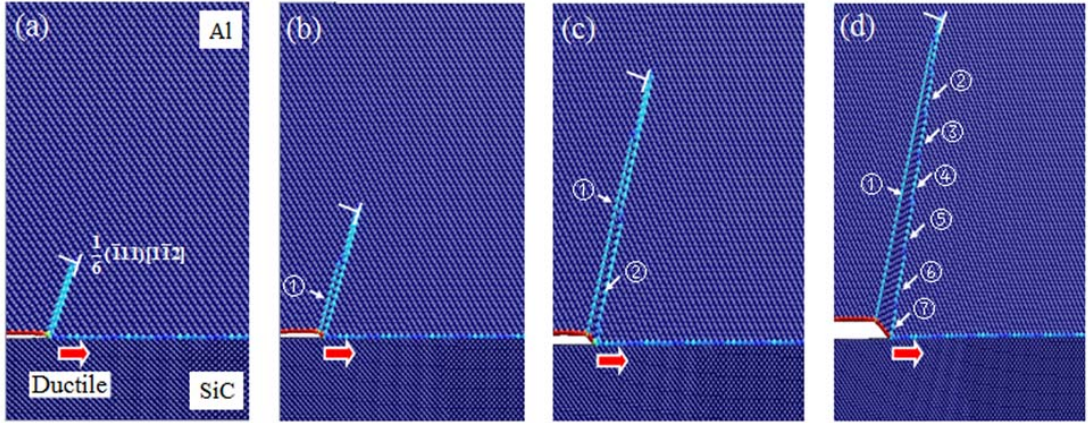

Fig. 4

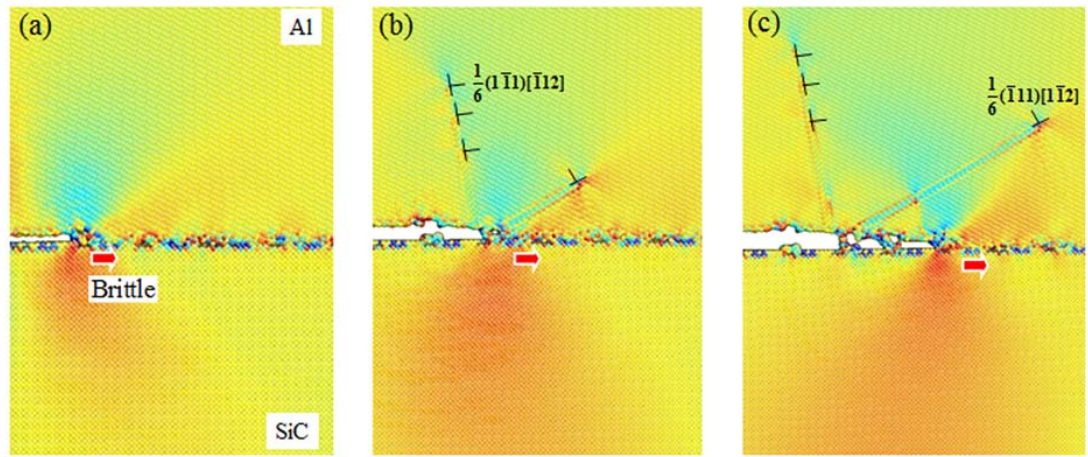

Fig. 5

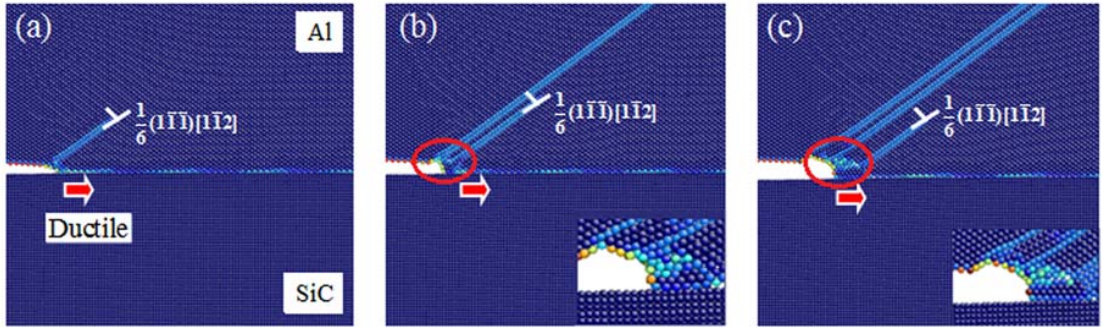

Fig. 6

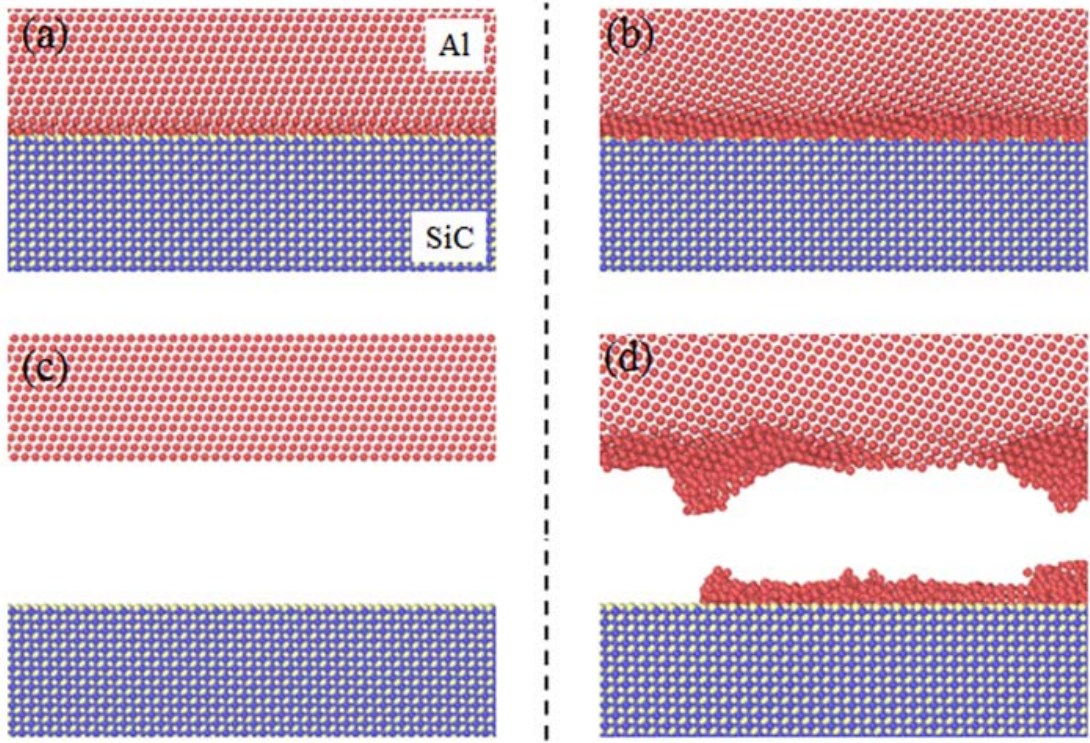

Fig. 7

Table 1 Al interface orientations, simulation cell dimensions (X|Y|Z), the number of atoms in the models with cracks

| Case | Al surfaces | Dimensions X×Y×Z (nm) | Atoms number |
|------|-------------|------------------------|--------------|
| A | [001] | 81.0×51.4×1.5 | 466853 |
| B | [1$\bar{1}$0] | 106.0×50.0×1.5 | 595296 |
| C | [1$\bar{1}\bar{2}$] | 84.1×51.5×1.5 | 485603 |
| D | [1$\bar{1}$1] | 99.2×44.6×1.5 | 497652 |
| E | [2$\bar{2}$1] | 85.9×53.4×1.5 | 514233 |
| F | [1$\bar{1}$4] | 116.6×37.8×1.5 | 493492 |
| G | [1$\bar{1}$3] | 95.0×53.6×1.5 | 570692 |
| H | [$\bar{3}$32] | 94.0×49.4×1.5 | 519195 |

Table 2. The actual fracture work $E_{act}$, the Griffith fracture work $E_{Theo}$ and the adhesive strength coefficient (ASC) $\lambda$, of different models.

| Case | $E_{act}$ (eV) | $E_{Theo}$ (eV) | $\lambda$ |
|------|----------------|-----------------|-----------|
| A    | 6953           | 8497            | 0.82      |
| B    | 8146           | 8162            | 1.00      |
| C    | 8287           | 9446            | 0.88      |
| D    | 8878           | 9132            | 0.97      |
| E    | 8353           | 9826            | 0.85      |
| F    | 8392           | 9630            | 0.87      |
| G    | 9231           | 11034           | 0.84      |
| H    | 8393           | 8888            | 0.94      |